\documentclass[12pt, preprint]{aastex}
\usepackage{epstopdf}
\usepackage{graphicx,epsfig}
\usepackage{natbib}
\usepackage{float}
\shorttitle{Identification of FPs}
\shortauthors{Abdul-Masih, Michael}

\begin{document}

\title{\emph{Kepler} Eclipsing Binary Stars. VIII. Identification of False Positive Eclipsing Binaries and Re-extraction of New Light Curves}

\author{
Michael Abdul-Masih\altaffilmark{1,2},
Andrej Pr\v sa\altaffilmark{1},
Kyle Conroy\altaffilmark{3},
Steven Bloemen\altaffilmark{4},
Tabetha Boyajian\altaffilmark{5},
Laurance R. Doyle\altaffilmark{6,7},
Cole Johnston\altaffilmark{1},
Veselin Kostov\altaffilmark{8},
David W. Latham\altaffilmark{9},
Gal Matijevi\v c\altaffilmark{1},
Avi Shporer\altaffilmark{10,11},
John Southworth\altaffilmark{12}
}

\altaffiltext{1}{Villanova University, Dept. of Astrophysics and Planetary Science, 800 E Lancaster Ave, Villanova, PA 19085}
\altaffiltext{2}{Department of Physics, Applied Physics and Astronomy, Rensselaer Polytechnic Institute, Troy, NY 12180}
\altaffiltext{3}{Vanderbilt University, Dept. of Physics and Astronomy, VU Station B 1807, Nashville, TN 37235}
\altaffiltext{4}{Department of Astrophysics/IMAPP, Radboud University Nijmegen, 6500 GL Nijmegen, The Netherlands}
\altaffiltext{5}{J.W. Gibbs Laboratory, Yale University, 260 Whitney Avenue, New Haven, CT 06511}
\altaffiltext{6}{IMoP at Principia College, Elsah, Illinois  62028}
\altaffiltext{7}{SETI Institute, 189 Bernardo Ave. Mountain View, California 94043}
\altaffiltext{8}{Department of Astronomy \& Astrophysics at U of T, Toronto, Ontario, Canada M5S 3H4}
\altaffiltext{9}{Harvard-Smithsonian Center for Astrophysics, 60 Garden Street, Cambridge, MA 02138}
\altaffiltext{10}{Jet Propulsion Laboratory, California Institute of Technology, 4800 Oak Grove Drive, Pasadena, CA 91109, USA}
\altaffiltext{11}{Sagan Fellow}
\altaffiltext{12}{Astrophysics Group, Keele University, Staffordshire, ST5 5BG, UK}

\begin{abstract}
The \emph{Kepler} Mission has provided unprecedented, nearly continuous photometric data of $\sim$200,000 objects in the $\sim$105 deg$^{2}$ field of view from the beginning of science operations in May of 2009 until the loss of the second reaction wheel in May of 2013. The \emph{Kepler} Eclipsing Binary Catalog contains information including but not limited to ephemerides, stellar parameters and analytical approximation fits for every known eclipsing binary system in the \emph{Kepler} Field of View. Using Target Pixel level data collected from \emph{Kepler} in conjunction with the \emph{Kepler} Eclipsing Binary Catalog, we identify false positives among eclipsing binaries, i.e. targets that are not eclipsing binaries themselves, but are instead contaminated by eclipsing binary sources nearby on the sky and show eclipsing binary signatures in their light curves. We present methods for identifying these false positives and for extracting new light curves for the true source of the observed binary signal. For each source, we extract three separate light curves for each quarter of available data by optimizing the signal-to-noise ratio, the relative percent eclipse depth and the flux eclipse depth.  We present 289 new eclipsing binaries in the \emph{Kepler} Field of View that were not targets for observation, and these have been added to the Catalog. An online version of this Catalog with downloadable content and visualization tools is maintained at \texttt{http://keplerEBs.villanova.edu}.
\end{abstract}

\section{Introduction}
The \emph{Kepler} mission was launched in 2009 and provided photometric data for $\sim$200,000 objects in the 105 deg$^{2}$ contained in the \emph{Kepler} Field of View (FOV) (Batalha et al., 2013). Each of the 95 million \emph{Kepler} pixels cover 3.98$\times$3.98 arc seconds and are designed to maximize the number of resolvable stars with magnitudes brighter than 15. Further details and specifications regarding the \emph{Kepler} mission can be found in Koch et al. (2010) and Borucki et al. (2010). There are approximately 500,000 objects in the \emph{Kepler} Field of View which are brighter than $V = 16$, however only $\sim$200,000 were assigned as targets for observation, leaving many bright objects in the field unobserved (Batalha et al., 2010). Since the main goal of \emph{Kepler} is to find Earth-sized planets in the habitable zone of Sun-like stars, the targets that were chosen for observation were those with the highest potential for terrestrial planet detection (Borucki et al., 2008). Thus, many objects in the \emph{Kepler} FOV have not been observed. Due to the proximity of some of these unobserved objects to the identified targets, the possibility of contaminated signals arises. Of the observed targets, 2772 eclipsing binaries have been found and cataloged in the \emph{Kepler} Eclipsing Binary Catalog, hereafter Catalog; (Kirk et al., 2015, accepted). Details regarding the identification and processing can be found in Pr\v sa et al. (2011), Slawson et al. (2011), Matijevi\v c et al. (2012) and Conroy et al. (2014).

The Target Pixel Files (TPFs) delivered by the \emph{Kepler} science office contain raw flux counts associated with each pixel as a function of time.  The light curve files returned by the \emph{Kepler} pipeline are generated by combining certain pixels in the TPF into an aperture to maximize the signal-to-noise ratio (SNR) (Bryson et al., 2010). These pixels are chosen based on their proximity to the target as well as the target's magnitude. Some objects are close enough to each other on the sky that the signal from one object can contaminate the pixels chosen for the aperture of the target. If the other object is a binary, then the automatically generated \emph{Kepler} integrated aperture light curve for the target would appear to have a binary signal even though it is not the source of that binary signal. This leads to a false positive binary signal.

Identification of these false positives and the re-extraction of new light curves for the true sources is essential to maintain the integrity and validity of the Catalog.  Binary systems are important because many of their parameters can be geometrically solved and modeled.  They can also be used to constrain evolutionary models as both stars in the binary system presumably formed at the same time.  Because of this, it is important that the intrinsic parameters measured by \emph{Kepler} belong to the source of the binary signal and not another nearby contaminated star.  Furthermore, by re-extracting new light curves, we can ensure that the binary signals obtained are as uncontaminated as possible.  There have been a few methods published concerning the identification of false positive.  Thompson et al. (2015) describes automated methods for identifying transit-like events with a specific focus on transiting exoplanets.  Coughlin et al. (2014) discusses ephemeris matching techniques, which can be used to identify cases where multiple \emph{Kepler} targets show the same transit signal (same or integer multiple period and same time of minimum).  Bryson et al. (2013) describes centroid analysis techniques to identify the location of transit sources.  In addition, several visualization techniques are discussed to identify false positives using TPF data.  These techniques are primarily designed for exoplanets, and therefore certain issues arise when trying to adapt them to binary systems.  The automated transit identification methods specifically remove non-exoplanet-like transit events, which presents a problem for systems such as overcontact binaries whose signal does not resemble an exoplanet transit.  Similarly, the ephemeris matching technique described in Coughlin et al. (2014) presents a problem for eclipsing binaries with similar primary and secondary eclipse depths because there is a possibility that the primary and secondary eclipses could be swapped for different objects. The method described in Coughlin et al. (2014) checks for period matches and consistent times of minimum but does not account for phase differences which would be seen if the primary and secondary were swapped. Exoplanet light curves do not show deep secondary eclipses, so this is rarely an issue for exoplanet ephemeris matching.  The methods described in Bryson et al. (2013) do not account for all background objects that are not in the \emph{Kepler} Input Catalog.  Furthermore, the visualization techniques described find a correlation between the signal from each pixel and a transit model which is designed to fit planetary transits and not eclipsing binaries.  For these reasons, we present a method that addresses these issues and is suited for non-planetary transit signals.

\section{Identification of False Positives}
For the purposes of this study, we define a \emph{Kepler} object as a system with a KIC designation, a \emph{Kepler} target as a \emph{Kepler} object that was chosen for observation and thus has been observed by \emph{Kepler}, and a background object as a system without a KIC designation.  A false positive is a case where the signal from a binary object in close proximity to a target contaminates the aperture pixels, causing the target light curve to show a binary signal. This means that the incorrect object is being identified as the source of the binary signal. There are two types of false positives that we observe: the first is target-target (i.e. two \emph{Kepler} targets chosen for observation) and the second is target-object (i.e. one \emph{Kepler} target chosen for observation and either one unobserved \emph{Kepler} object or one background object). In addition, there is a possibility of a combination of these two where multiple targets are contaminated by the same true eclipsing binary source.

An example of a target-target false positive can be seen with KIC 5467102 and KIC 5467113, which are separated by less than 10 arc seconds (Fig. 1).  The light curves for both objects show a binary signal with the same period and phase. Due to the proximity of these two targets, the TPFs associated with each target share a number of pixels. Even when apertures do not overlap, signal from one target can still contaminate that of the other, if one is much brighter, so its point spread function (PSF) spills over its own aperture. Fig. 1 shows the raw and detrended light curves and TPFs for each object.  For visualization purposes, the shared pixels are outlined in red. Comparing the percent eclipse depths of the two objects shows that the target responsible for the binary signal is KIC 5467113 as the percent eclipse depth is much larger. 

\begin{figure}[H]
\includegraphics[width=0.5\textwidth]{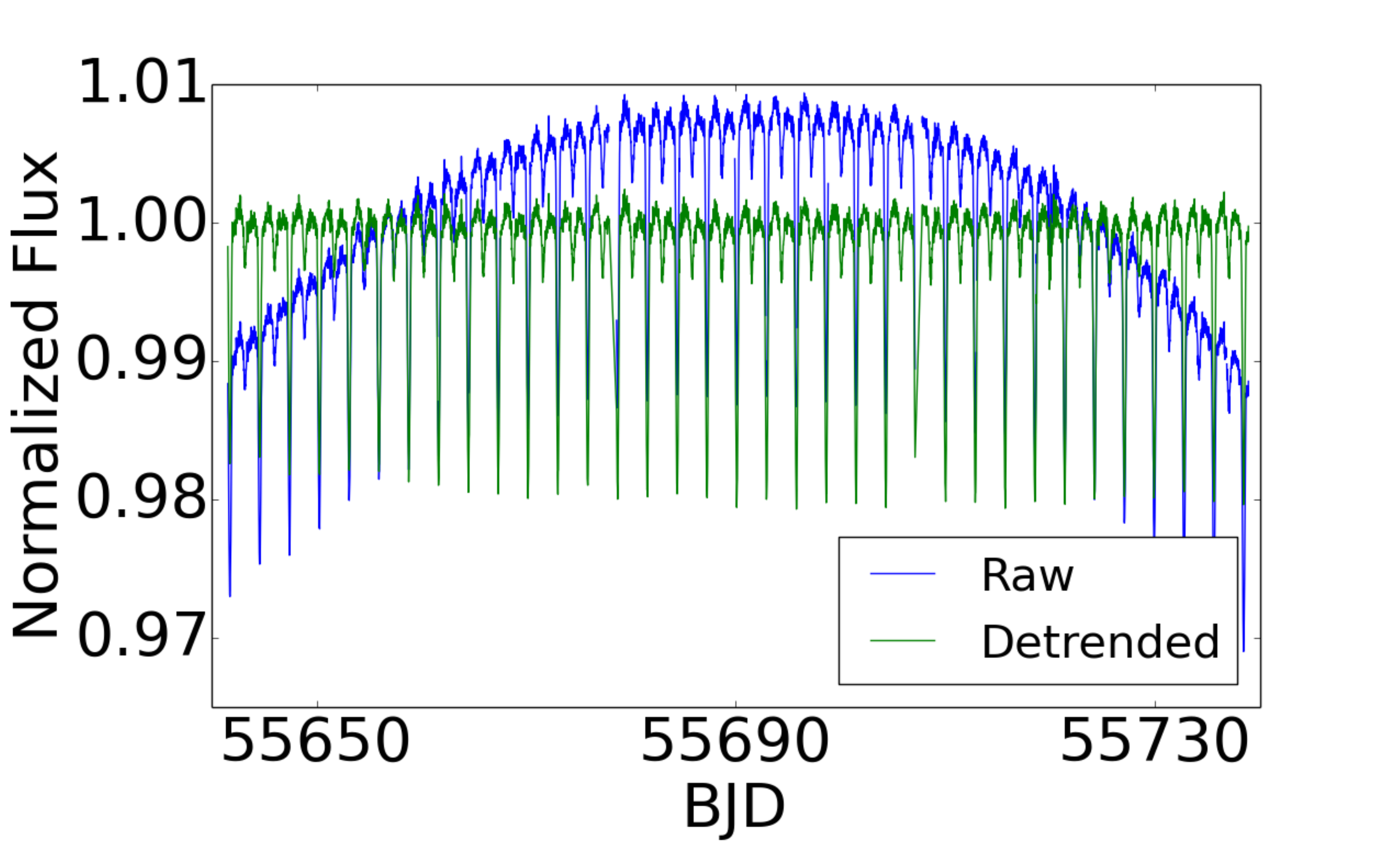}
\includegraphics[width=0.5\textwidth]{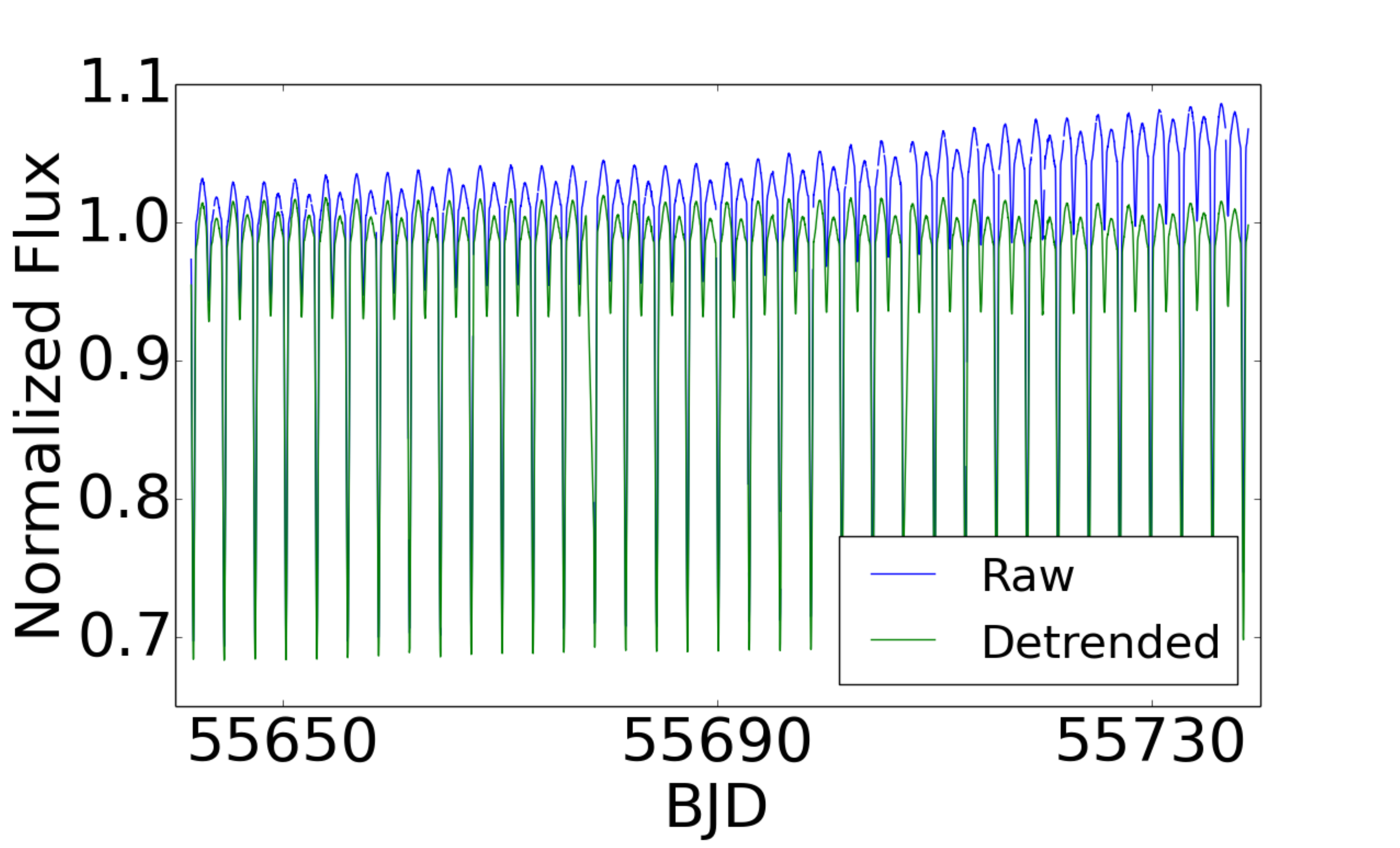} \\
\includegraphics[width=0.5\textwidth]{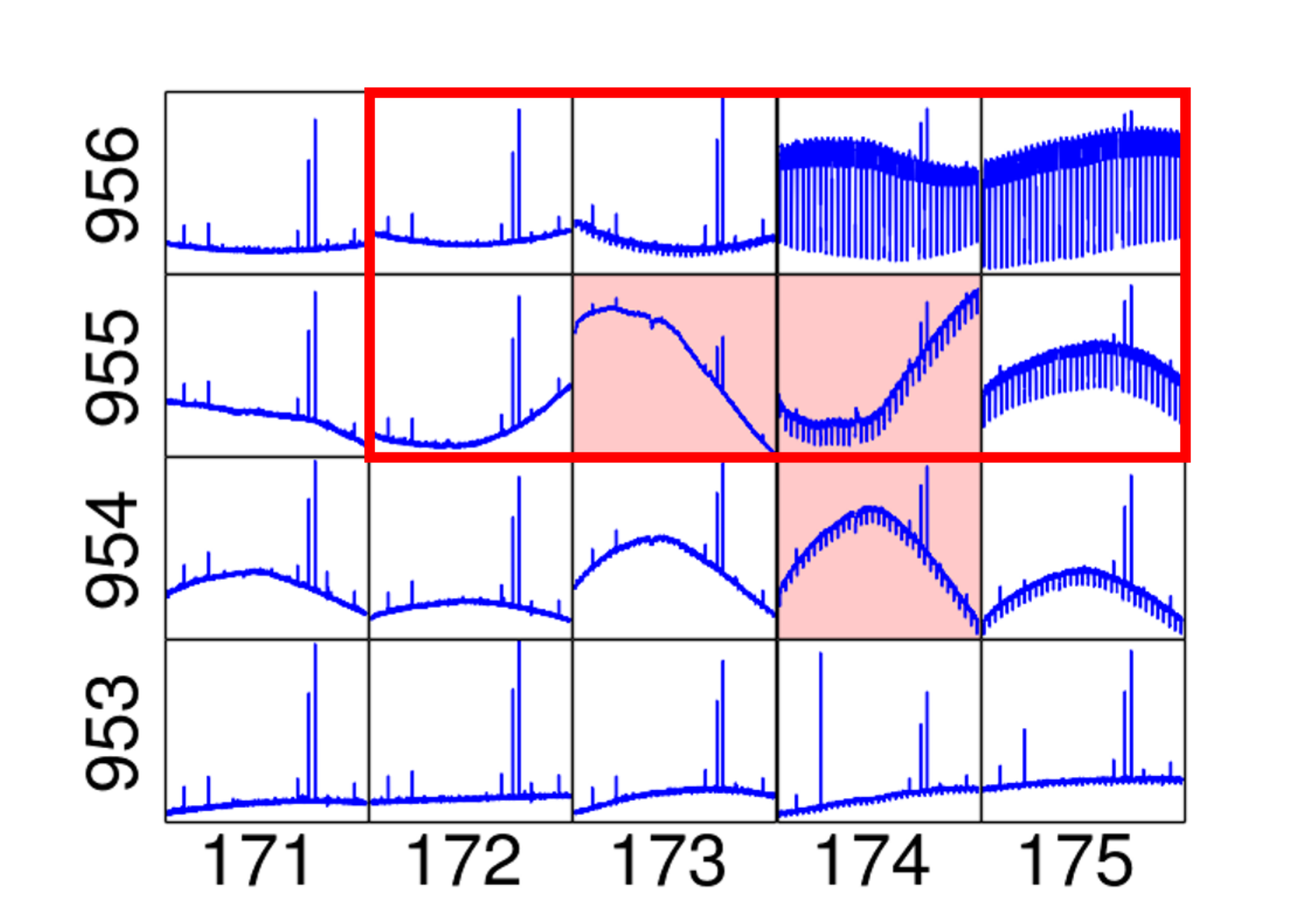}
\includegraphics[width=0.5\textwidth]{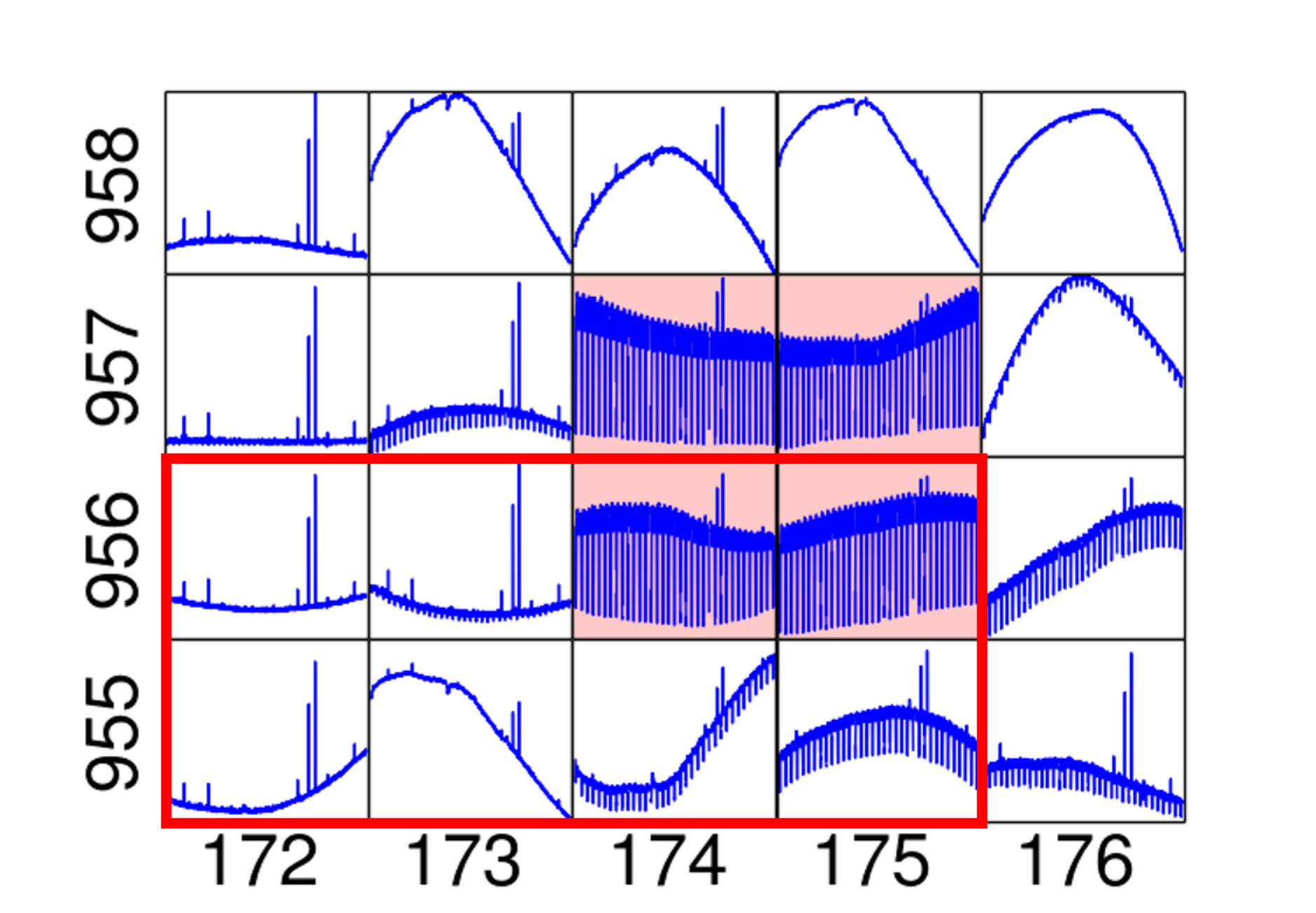} \\
\caption{
\label{fig:target-target}
Top left and right: the raw (blue) and detrended (green) Quarter 9 light curves for KIC 5467102 and KIC 5467113, respectively. Bottom left and right: the TPFs showing raw data for Quarter 9 for KIC 5467102 and KIC 5467113. The outlined pixels represent the overlap between the two windows while the pixels with a red background represent the pixels that were chosen for the optimal aperture for each.  The pixel row and column numbers are indicated.
}
\end{figure}

Target-object false positives are more difficult to interpret than target-target false positives because analyzing the light curves is insufficient to determine whether or not the target in question is a false positive. Since non-target objects do not have light curves to compare to, raw TPF data are required.  By analyzing the raw TPF data, one can determine which pixels in the TPF contain the binary signal in question and where this signal is most likely originating. An example of such a case can be seen in KIC 4356766, shown in Fig. 2.  The light curve shows a binary signal, however when the TPF data are analyzed, it is obvious that KIC 4356766 is not responsible for the binary signal.  The bottom plot in Fig. 2 shows the TPF data for each pixel.  The pixels with the strongest signals are to the upper right of the TPF indicating that the binary producing this signal is not KIC 4356766 and that the signal from the actual binary is bleeding into the aperture for KIC 4356766.  For this reason, the integrated aperture light curves can be misleading if the TPF data are not analyzed in conjunction with the light curve data.  KIC 4356766 is used as an example throughout the paper to demonstrate various analysis techniques.

\begin{figure}[H]
\center
\includegraphics[width=0.5\textwidth]{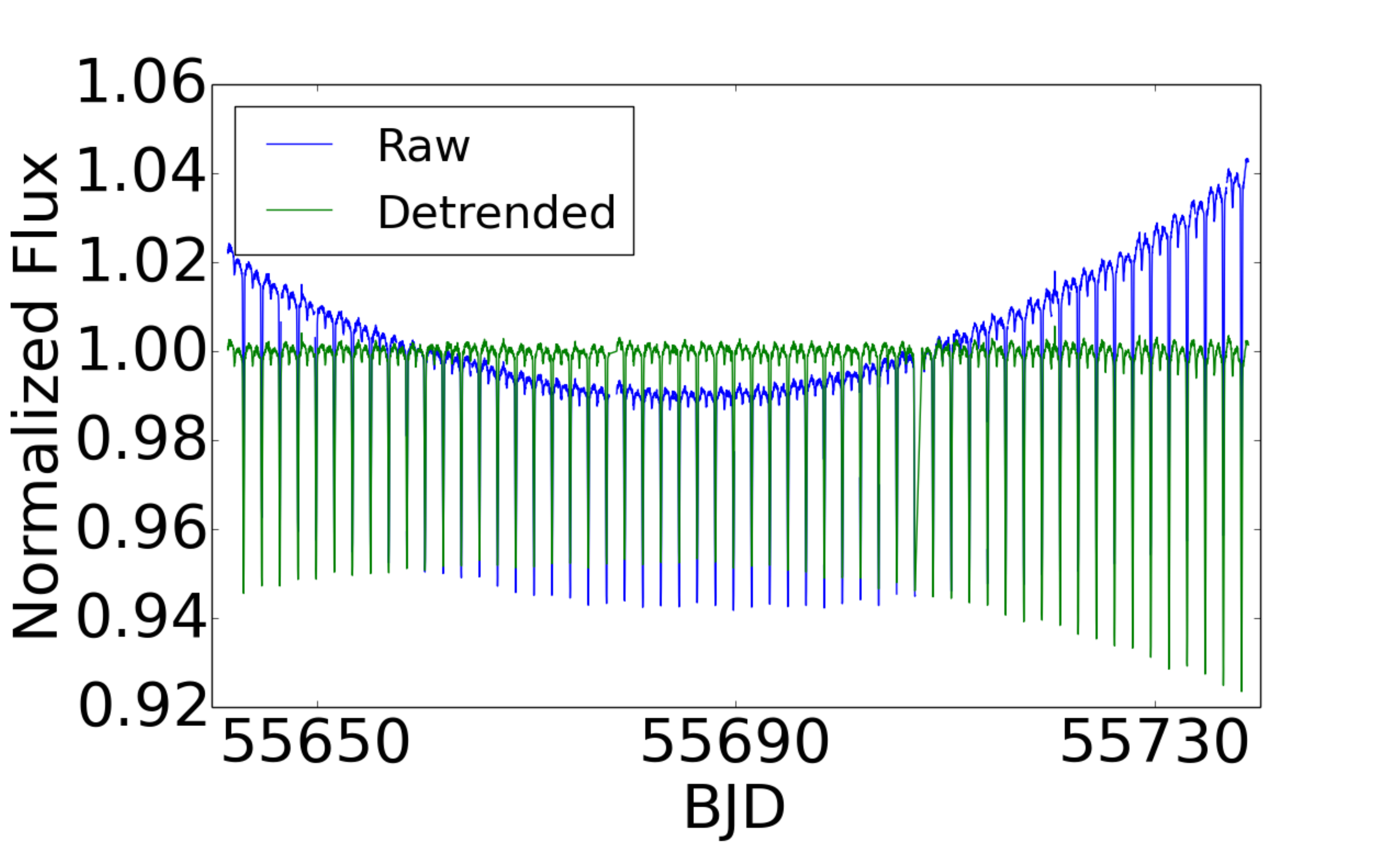}
\includegraphics[width=\textwidth]{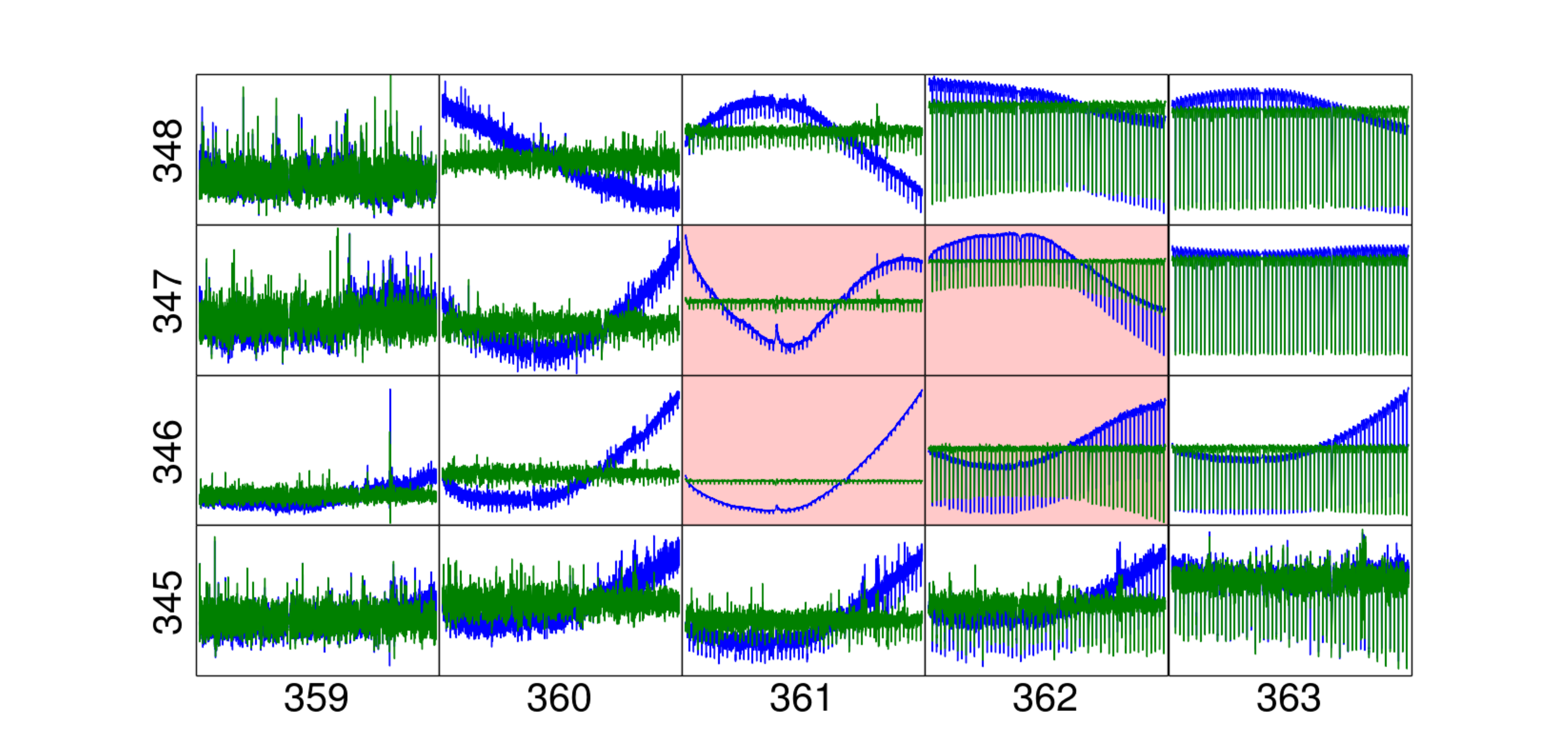}
\caption{
\label{fig:4356766_lc}
Top: Quarter 9 \emph{Kepler} light curve for KIC 4356766. The blue represents the raw light curve and the green is the detrended light curve. Bottom: raw (blue) and detrended (green) data for each pixel in the TPF for Quarter 9 of KIC 4356766 with pixels in \emph{Kepler} aperture indicated with a pink background. The pixel row and column numbers are provided on the axes.
}
\end{figure}

Target-target false positive candidates can be identified through ephemeris matching.  By comparing the ephemerides as well as the shapes of the phased light curves, we can determine whether the binary signals seen in two targets likely originate from the same source.  If the ephemerides and light curve shapes are consistent, then we look at the angular separation of the two objects to see if they are close enough for direct signal contamination. For a majority of cases, the targets are within 20 arc seconds of each other, however for brighter binaries ($V < 10$), the signal can contaminate pixels further away.  The target-target cases where the source is a bright binary are easier to detect because they often contaminate several targets in the area.  Few such cases have been observed, however, and they make up a small percentage of the false positives identified. If the targets have a larger separation, then the cause of contamination might be optical cross-talk.  Cross-talk occurs due to the optics of the telescope, which can cause some light to be scattered to other sections of the CCD or a different CCD  altogether (Caldwell et al. 2010). If a binary signal is scattered, it can appear in pixels very far from the true source. Evidence of cross-talk can be seen in the TPFs, as the signal tends to be multiplicative in nature and thus shows up in all pixels with high flux in the contaminated area.

The method for identifying false positives that we introduce here involves several tiers: detrending the signal in each pixel (removing any data points with bad quality flags and then fitting polynomials to the signal to remove trends over the quarter), comparing the phase-folded detrended signal of each pixel to the light curve (pixel-LC) and neighboring pixels (pixel-pixel), comparing the signal-to-noise ratio of each pixel to the signal-to-noise ratio of the integrated aperture light curve, analyzing centroid movement as a function of time and phase and finally comparing the location where the signal is suspected to originate from with the locations of known \emph{Kepler} and background objects. All of these steps require examination of the TPF files. Comparing the pixel light curves to the integrated aperture light curve (pixel-LC correlation) and comparing the pixel light curves to the neighboring pixel light curves (pixel-pixel correlation)  are discussed in Sections 2.1 and 2.2, respectively.  These two steps utilize the Pearson correlation coefficient (PCC), which is a measure of the linear correlation between two variables, to compare the phase-folded detrended signals in question. The PCC can have a value between -1 and 1 where 1 is perfect correlation and -1 is perfect anti-correlation, with 0 being no correlation. Comparing the SNR of each pixel in the TPF to the scaled SNR of the light curve, as well as examining the centroid data and comparing these to a UKIRT image of the area, is discussed in Section 2.3.

\subsection{Pixel-LC Correlation}
The pixel-LC correlation tier is useful for determining which pixels contain the signal that is seen in the integrated aperture light curve. This is accomplished by first detrending both the pixel and the integrated aperture light curves. The PCC is then determined between each pixel and the integrated aperture light curve.  The left plot in Fig. 3 shows the pixel-LC comparison plot for KIC 4356766. The PCC values are plotted as a heat map with redder colors representing higher correlation (typically $>0.9$) and bluer colors representing lower correlation values (typically $<0.3$). The WCS coordinates provided in the TPF file are used to plot this heat map over sky coordinates. The locations of the \emph{Kepler} target in question as well as other \emph{Kepler} targets and \emph{Kepler} objects are overplotted in green and red, respectively. The size of the circle represents the \emph{Kepler} magnitude of each object. This plot shows where in the TPF the signal is localized as well as which \emph{Kepler} objects are nearby. 

\subsection{Pixel-Pixel Correlation}
The pixel-pixel correlation tier is useful for determining where on the TPF unique non-instrumental signals are centered. As with the pixel-LC correlation, all of the pixel light curves are first detrended.  The PCC is then determined between each pixel and each of the pixels directly surrounding it. For each pixel, the PCCs between itself and its neighbors are averaged and this value is assigned to the pixel. The right plot in Fig. 3 shows the pixel-pixel correlation plot for KIC 4356766. The PCC values are plotted as a heat map and the \emph{Kepler} objects are overplotted as with the pixel-LC correlation. With the pixel-pixel correlation, the heat map shows the areas of most similarity in the TPF, which effectively probes the PSFs of all sources in the TPF. 

\begin{figure}[H]
\center
\includegraphics[width=0.45\textwidth]{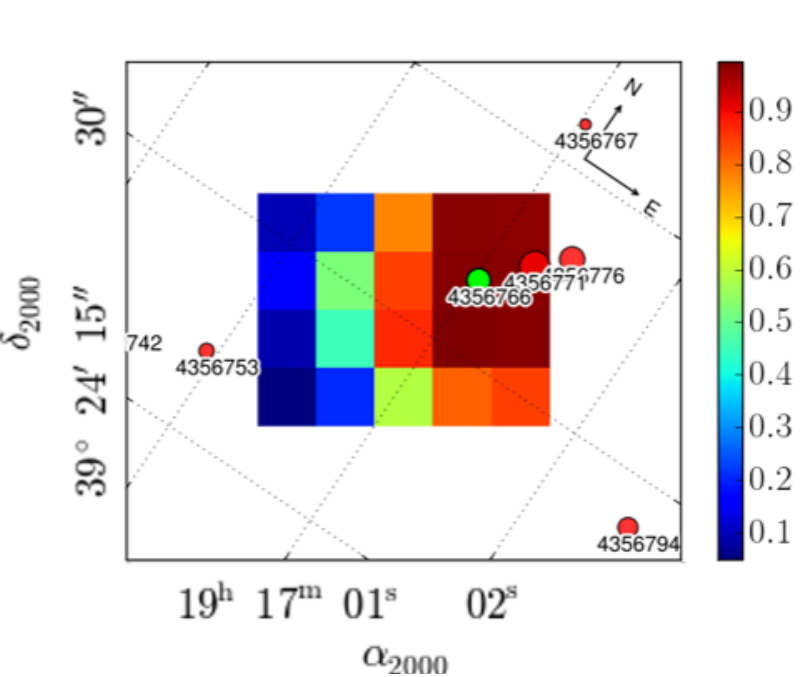}
\includegraphics[width=0.45\textwidth]{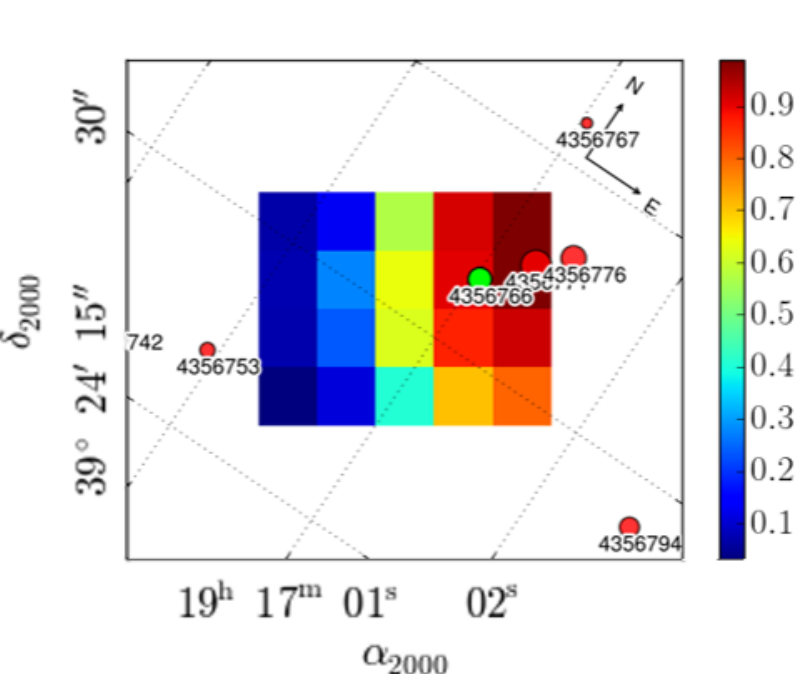}
\caption{
\label{fig:4356766_lc}
Left and right: pixel-LC correlation plot and pixel-pixel correlation plot, respectively.  The color scale represents the pixel correlation value as determined by each method and the location of the \emph{Kepler} objects are overplotted.  The size of the plotted objects is representative of the \emph{Kepler} magnitude and the green point is the target in question.
}
\end{figure}

\subsection{SNR Comparison and Centroid}
The SNR comparison and centroid plot tier is comprised of three steps.  The first step is similar to the pixel-LC and pixel-pixel correlation tiers, however the heat map is calculated by comparing the SNR of each pixel to the SNR of the light curve. The second step computes the location of the centroid throughout the quarter. The third step involves analyzing the locations of known \emph{Kepler} targets, \emph{Kepler} objects and background objects and comparing them to the heat maps. 

The first step is set up in the same way as the pixel-pixel and pixel-LC correlation plots except that the heat map is computed by comparing the SNR of each pixel to the scaled SNR of the integrated aperture light curve. The color scale goes from red, which represents stronger binary signal than the average pixel in the optimal aperture of the integrated aperture light curve, to white, which represents signal comparable to the average pixel in the aperture of the integrated aperture light curve, to blue, which represents signal weaker than the average pixel in the aperture of the integrated aperture light curve. In addition, the \emph{Kepler} defined aperture is outlined in black and the positions of the centroid points are plotted in white. 

To calculate the heat map, the primary eclipse depths are determined by first detrending and normalizing the raw flux data from each pixel and then phase-folding them over the period of the binary signal detected. The flux change between in- and out-of-eclipse is determined by polyfits, a method that fits a piecewise chain of $n$th order polynomials to the phase-folded light curve and comparing the in-eclipse and out-of-eclipse regions (Pr\v sa et al. 2008). This primary eclipse depth is then divided by the standard deviation of the out-of-eclipse flux to give the SNR for each pixel. This value is then compared to the SNR of the light curve which is calculated in the same way. Each pixel's SNR is divided by the SNR of the light curve and this ratio is multiplied by a scaling factor. The scaling factor is determined by first dividing the out-of-eclipse baseline of the light curve by the number of pixels in the aperture to find the average pixel contribution. Each pixel's baseline is then divided by the average pixel contribution and then this ratio is square-rooted, giving the scaling factor. The SNR ratios multiplied by the scaling factor give the values for each pixel in the heat map (Eq. 1), which is plotted on a log-scale for visualization purposes. In Eq. 1, $b_{pixel}$ and $b_{LC}$ are the baseline for the pixel and light curve, respectively and n is the number of pixels in the optimal aperture. This heat map shows where the binary signal is most prominent. 

\begin{equation}
\textrm{SNR comparison heat map value = $\frac{\mathrm{SNR}_{pixel}}{\mathrm{SNR}_{LC}} \sqrt{\frac{b_{pixel}}{\left(\frac{b_{LC}}{n}\right)}}$}
\end{equation}

The second step is the zoomed-in temporal and spatial progression of the centroid throughout the quarter.  The color outline of each point represents the time through the quarter starting with blue in the beginning of the quarter and ending with red. The face color of the symbol in grayscale represents the detrended flux at each data point. Darker shades represent lower flux.  As the binary eclipses, the flux drops and the centroid is pulled away from the binary towards other light sources. The top left and right panels of Fig. 4 show the SNR comparison heat map and centroid plots for Quarter 9 of KIC 4356766, respectively.

\begin{figure}[h]
\includegraphics[width=\textwidth]{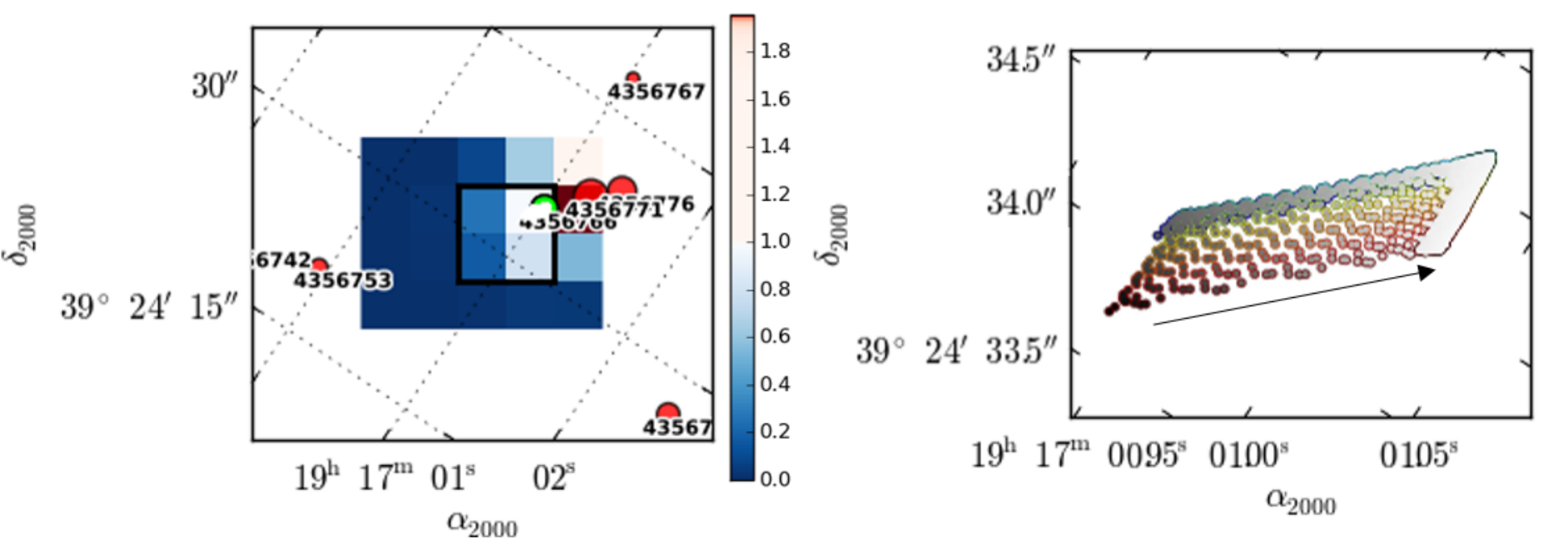}
\center
\includegraphics[width=0.25\textwidth]{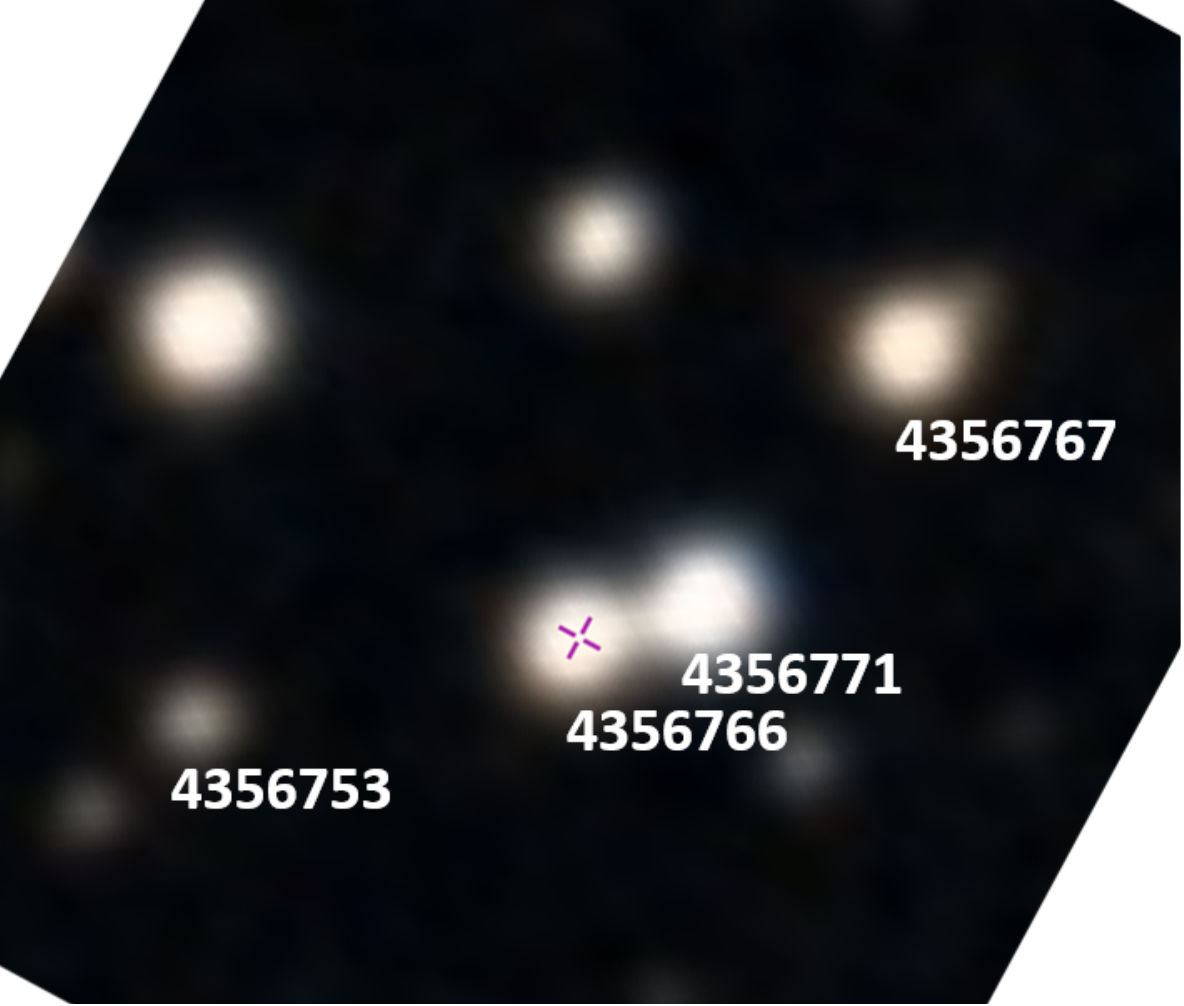}
\caption{
\label{fig:heat_map_centroid}
Top: the SNR comparison heat map and centroid plots for Quarter 9 of KIC 4356766 with an arrow indicating the direction of the binary overplotted on the centroid plot.  Bottom: a UKIRT image of the sky centered on the target. \emph{Kepler} objects are labeled with their respective KIC designations.
}
\end{figure}

As the binary begins to eclipse, the centroid shifts away from it because it is contributing less light than when out-of-eclipse. Thus, by following the motion of the centroid while noting its position on the heat map, the direction towards the true source can be established. This, in conjunction with the SNR correlation plot, provides a powerful diagnostic of where the binary is located. By overplotting the locations and brightnesses of the \emph{Kepler} objects in the area, we can quickly determine which objects may be responsible for the signal.  If the signal does not appear to be coming from a \emph{Kepler} object, a UKIRT image of the area is used which includes objects that were not given a KIC designation. The bottom panel in Fig. 4 shows an example UKIRT image of the area around KIC 4356766.  From Figures 2, 3 and 4, it is evident that KIC 4356766 is a false positive being contaminated by KIC 4356771.  This tier is also useful for identifying cases of cross-talk since cross-talk manifests itself with the same binary signal appearing in each source in the TPF.  Fig. 5 shows an example of cross-talk.
 
\begin{figure}[h]
\center
\includegraphics[width=0.5\textwidth]{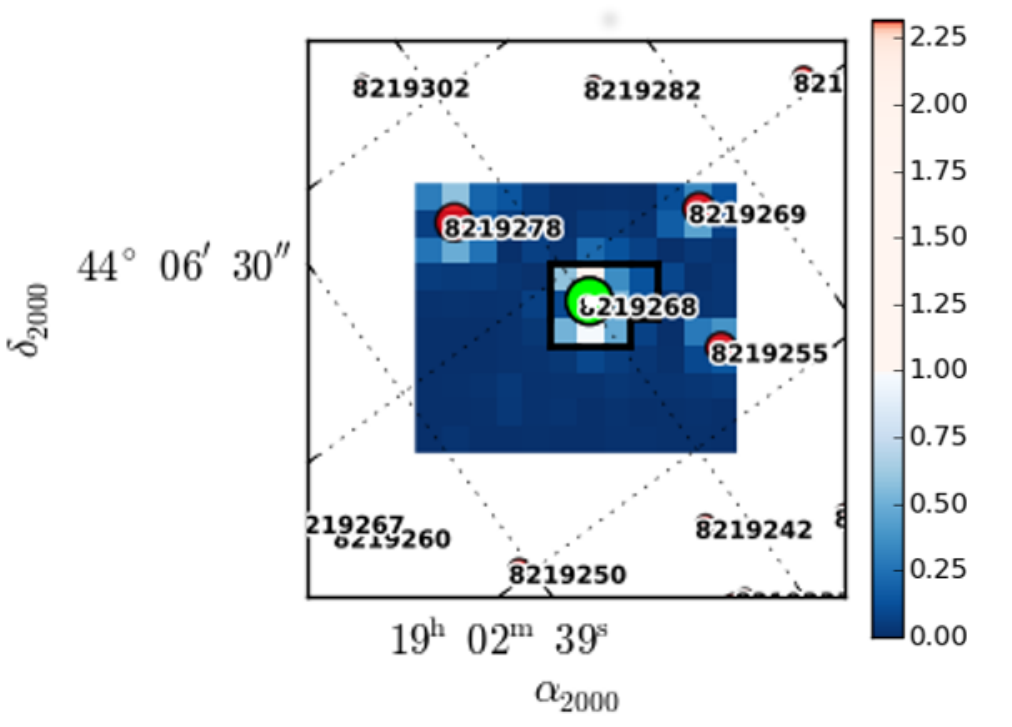}
\caption{
\label{fig:crosstalk}
SNR comparison heat map plot for Quarter 0 of KIC 8219268 showing the manifestation of cross-talk.
}
\end{figure}

\section{Re-extraction of New Light Curves}
Once the source of the binary signal is identified, a new light curve can be extracted by assigning a custom aperture in the TPF. Re-extraction is only done for target-object false positives because both members of a target-target false positive pair already have light curves associated with them. Using the false positive TPF, the re-extraction process is comprised of several automated steps and results in three new light curves generated for the true source of the binary signal for each quarter of available data.  These light curves are generated by optimizing the signal-to-noise ratio (SNR light curve), percent eclipse depth (PED light curve), and flux eclipse depth (FED light curve). The SNR light curve is meant to minimize the noise and be most similar to the original integrated aperture light curve.  The PED light curve is meant to give the most accurate percent eclipse depth, and the FED light curve is meant to capture all pixels with non negligible contributions to the total flux eclipse depth of the TPF.

Before re-extracting the new light curves, we obtain the SNR, the percent eclipse depth and the flux eclipse depth for the original false positive integrated aperture light curve.  The flux eclipse depth can be obtained by measuring the difference between the out-of-eclipse flux and the in-eclipse flux.  These flux values are determined by first fitting polyfits to the in-eclipse and out-of-eclipse regions, and then using the returned statistics which include the in-eclipse and out-of-eclipse boundaries (Pr\v sa et al. 2008).  The in-eclipse flux is determined by taking the median of the flux values of the middle 20\% of the in-eclipse region surrounding zero phase, and the out-of-eclipse flux is determined by taking the median of the flux values of the out-of-eclipse region.  The percent eclipse depth is then determined by dividing the flux eclipse depth by the out-of-eclipse flux. The signal-to-noise ratio is determined by first subtracting the polyfit solution from the phase-folded light curve. The noise is determined by calculating the standard deviation of the out of eclipse region and then the flux eclipse depth is divided by the noise to obtain the signal-to-noise ratio. 

The same calculations for the SNR, percent eclipse depth and flux eclipse depth are carried out for each pixel in the TPF window. The next step is to calculate the flux eclipse depth (FED) light curve, which contains at least 99\% of the total eclipsing binary signal seen in the TPF.  To do this we first calculate the total flux eclipse depth of the binary signal by adding the flux eclipse depths of each pixel in the TPF.  Starting with the pixel with the largest flux eclipse depth, we progressively add an additional pixel with the next largest flux eclipse depth until the new flux eclipse depth is greater than 99\% of the total flux eclipse depth of the TPF.  The pixels that were chosen are combined into a new mask.  Since the FED light curve is meant to capture all pixels with eclipse signals that contribute to the total flux eclipse depth of the TPF, no limit is placed on the number of pixels allowed in the aperture. The pixels chosen are not required to be adjacent to one another but typically they are adjacent.  Cases where the pixels in the aperture are not adjacent include low SNR binaries and optical cross-talk events.

The number of pixels contained in the optimal aperture of the integrated aperture light curve depends primarily on the \emph{Kepler} magnitude of the target.  Fig. 6 shows the number of pixels in the optimal \emph{Kepler} aperture as a function of magnitude for all objects in the Catalog.  We fit an inverse function to the curve and use it to approximate how many pixels should be in the SNR and percent eclipse depth (PED) apertures. Using the spread around the inverse fit, we put an upper limit on the number of pixels allowed in the final aperture.  For objects whose \emph{Kepler} magnitude is between $\sim10$ and $\sim18$, a good approximation for the upper limit seems to be 1.5 times the inverse fit, shown in Fig. 6. Due to the optics of the telescope, objects in different regions of the FOV will be focused differently and will require different numbers of pixels in their optimal apertures.  Objects closer to the center of the FOV will be more focused and will have a smaller PSF than objects towards the edges.  Using the approximation ensures that objects on the edges of the FOV are allowed enough pixels to account for the differences in PSF sizes. Since this approximation begins to fall off at fainter magnitudes, we set the minimum upper limit to be 4 pixels.  Objects with magnitudes $\mathrm{Kp} < 10$ are limited by the number of pixels in the TPF of the contaminated false positive. This upper limit is then compared to the number of pixels in the FED light curve aperture to further constrain the upper limit of allowed pixels.  This is done to remain consistent with the aperture sizes determined by \emph{Kepler}. For this reason, only the upper limit of the number of pixels for unobserved \emph{Kepler} objects are determined this way because they have a \emph{Kepler} magnitude associated with them. The upper limit for background objects is determined exclusively from the number of pixels in the FED light curve aperture.

\begin{figure}[h]
\center
\includegraphics[width=0.45\textwidth]{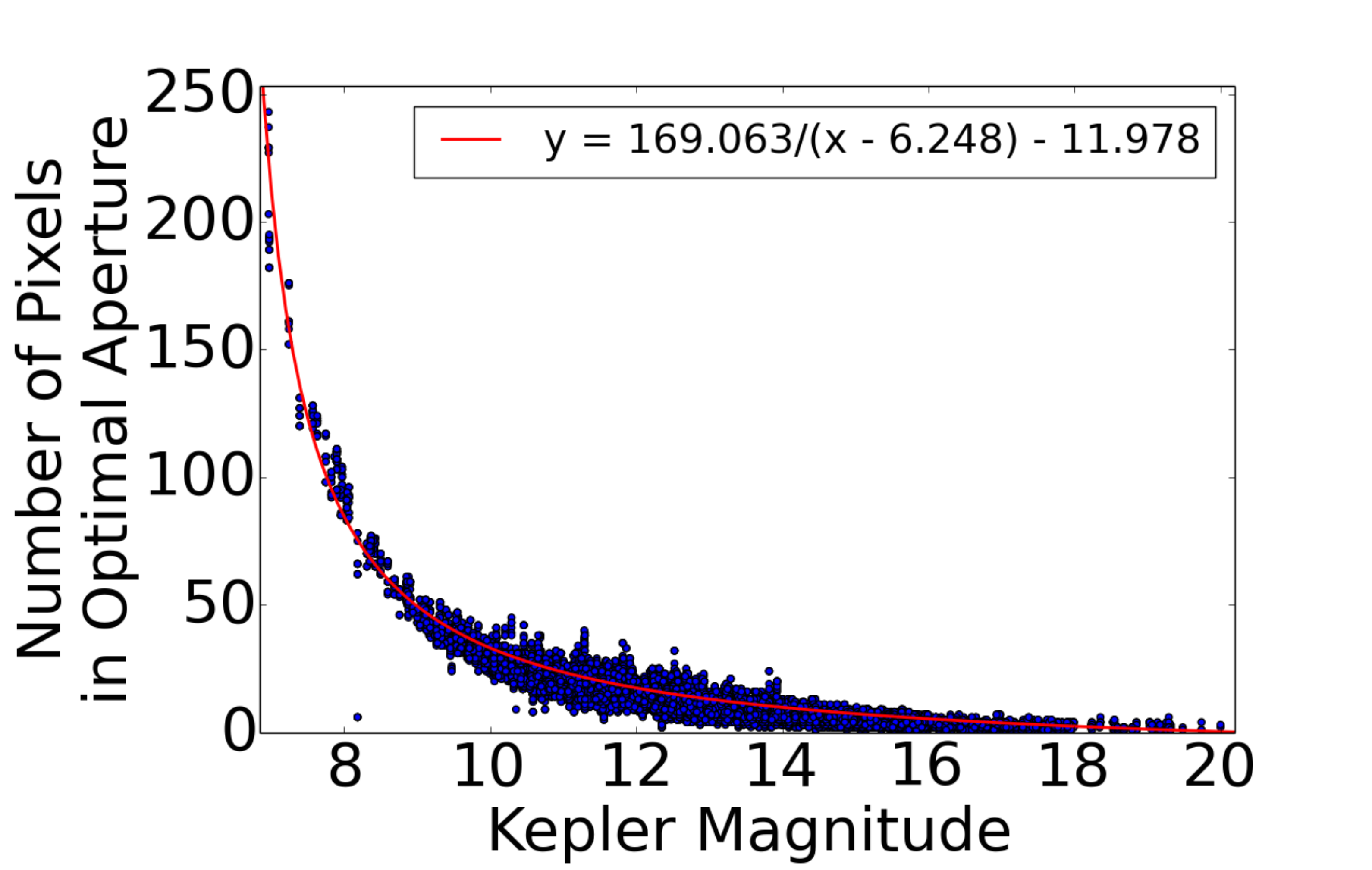}
\includegraphics[width=0.45\textwidth]{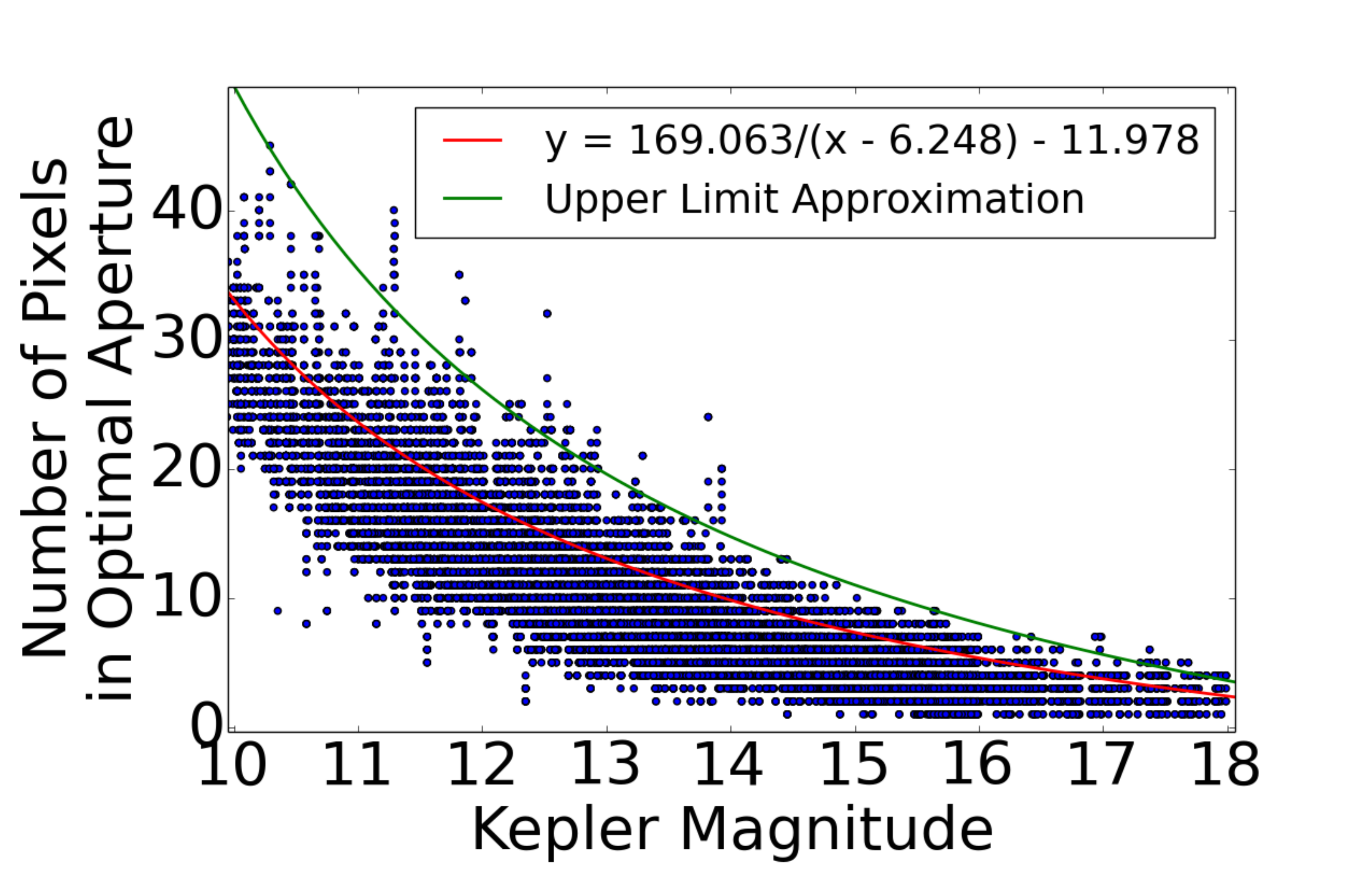}
\caption{
\label{fig:pix_mag}
Left: number of pixels chosen by \emph{Kepler} for the optimal aperture as a function of magnitude with best fit function overplotted in red. Right: zoomed-in portion of left plot showing the upper limit approximation in green and the best fit in red.
}
\end{figure}

We then calculate the new SNR light curve.  Starting with the two pixels with the highest SNRs, a new light curve is generated and the SNR is recorded. An additional pixel is added and the SNR of the new light curve is recalculated and recorded.  This process is repeated until the upper limit of the number of pixels allowed in the aperture is reached.  The pixels in the aperture that result in the highest SNR are then combined to form the SNR light curve. Once the SNR light curve is finalized, the same number of pixels used to generate the SNR light curve mask is used to generate the PED light curve mask. The pixels with the highest percent eclipse depth are chosen and combined to form the PED light curve.

\begin{figure}[h!]
\center
\includegraphics[width=0.5\textwidth]{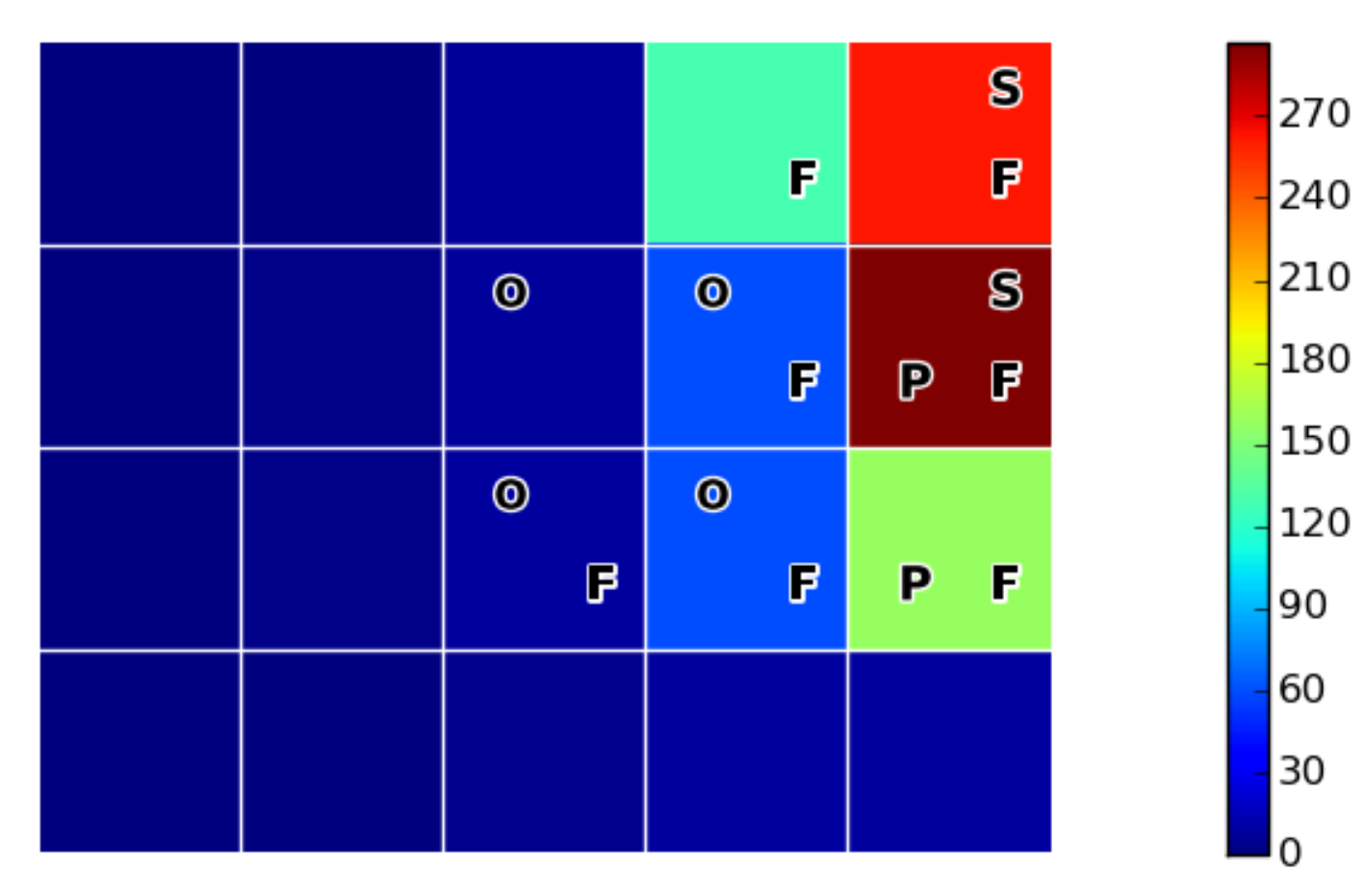} \\
\includegraphics[width=0.45\textwidth]{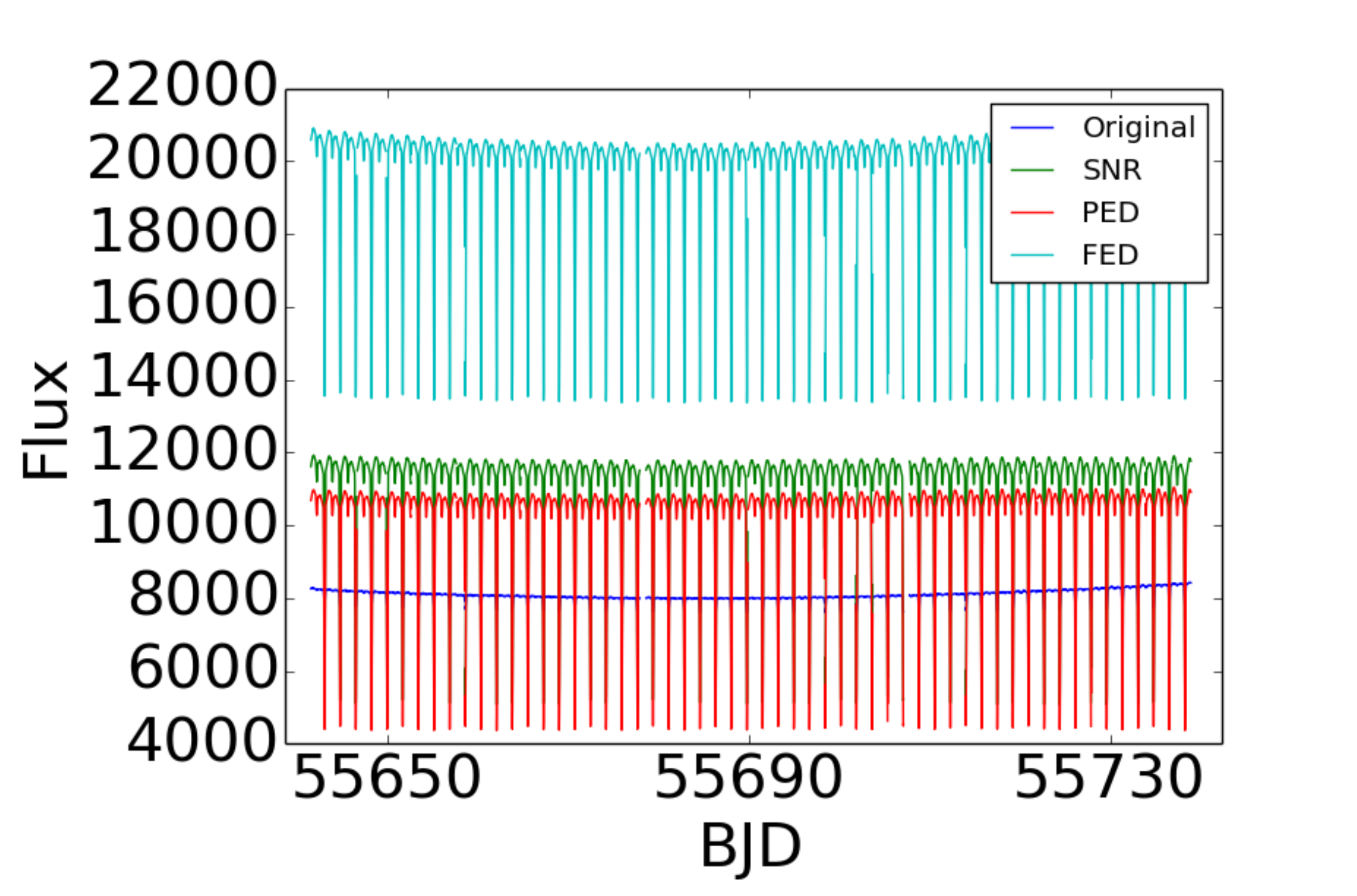}
\includegraphics[width=0.45\textwidth]{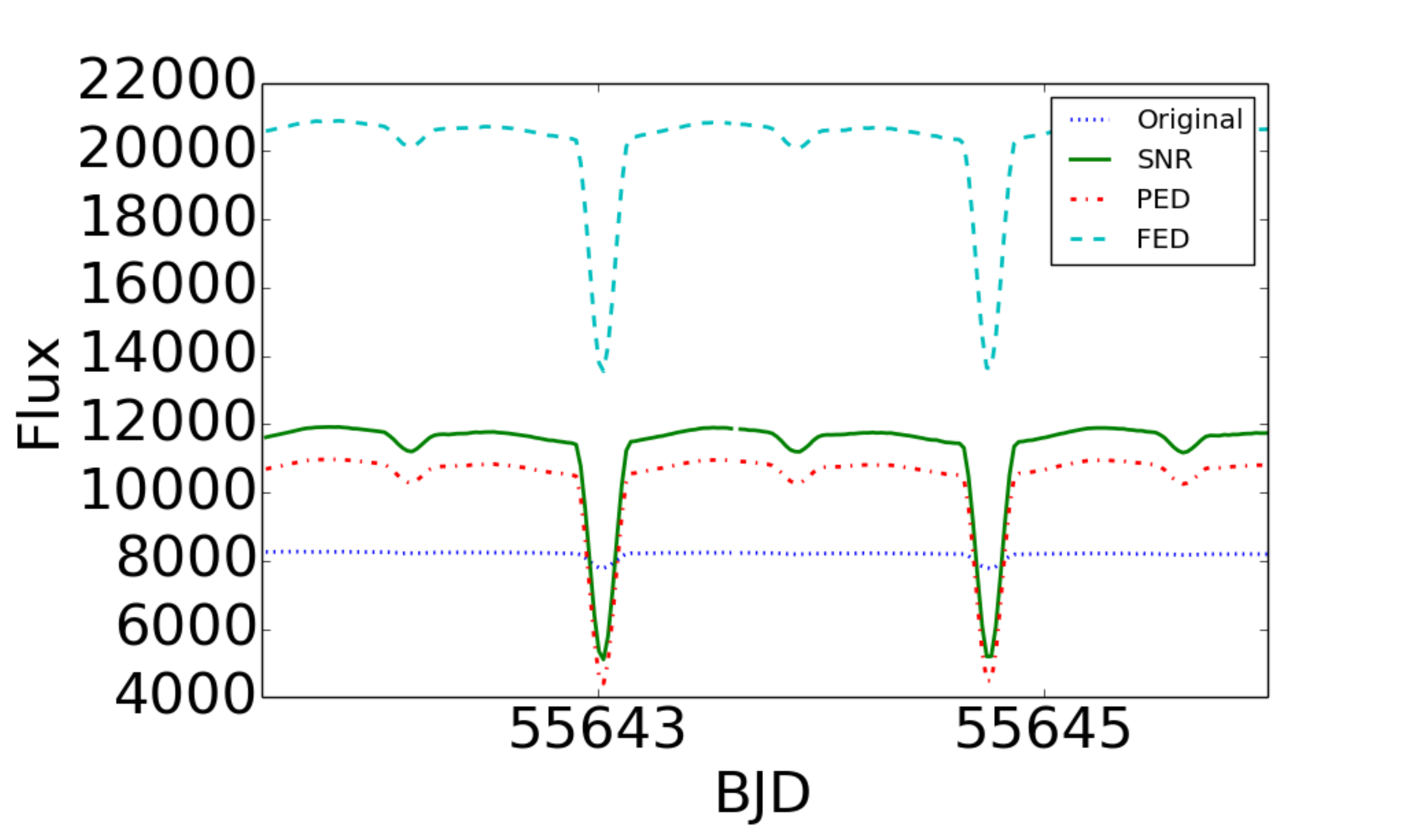}
\caption{
\label{fig:4356766_reextract}
Top: a heat map of the SNRs of each pixel from blue (lowest) to red (highest) in the TPF window for Quarter 9 of KIC 4356766.  The pixels in the original aperture are indicated with ``O'', the SNR aperture with ``S'', the PED aperture with ``P'' and the FED aperture with ``F''.  Bottom: two different views of the re-extracted light curves compared to the original light curve.  The left shows the whole quarter while the right shows a zoomed in portion of the beginning of the quarter.
}
\end{figure}

\section{Results}
A final light curve file is computed for each quarter of available data that contains the original light curve, as well as the three re-extracted light curves.  The original light curve file is kept in its entirety and the re-extracted light curve FITS tables, as well as their aperture information FITS Tables, are appended. Thus, the new re-extracted light curve file contains nine FITS tables in the following order: the original three tables from the integrated aperture light curve file describing the general information, the light curve data and the aperture information, respectively, the SNR light curve data and corresponding aperture information, the PED light curve data and corresponding aperture information, and the FED light curve data and its corresponding aperture information. These files are provided on the Catalog website.  The top plot in Fig. 7 shows different apertures chosen for Quarter 9 of KIC 4356766. The bottom two panels compare the new light curves to the old.

The pixel-pixel correlation, pixel-LC correlation, SNR correlation and centroid tiers are used to remove all false positive eclipsing binaries in the Catalog, the details of which can be found in Kirk et al. (2015, accepted).  In addition, the shapes of the phase-folded light curves and the ephemerides of every object in the Catalog were compared, and objects with periods within 1\% of each other were checked manually for consistency. Of the 624 false positives identified, 289 were found to be target-object false positives and new light curves were re-extracted for 285 of these false positives. The remaining four could not be re-extracted because their periods were longer than one quarter. Polyfit requires a full period of data to fit the polynomials, so if the period is longer than $\sim90$ days, the length of a quarter, then polyfit will fail. Of the 285, 163 have a KIC designation, while the remaining 122 are background objects without KIC designations.  Comparing the SNRs of the original integrated aperture light curves and the new re-extracted SNR light curves shows a significant increase in SNR (Fig. 8).  The top panel of Fig. 8 shows a plot of the SNR of the newly generated SNR light curves as a function of the SNR of the corresponding integrated aperture light curves. Each object is represented by a single point which corresponds to the quarter with the SNR light curve with the highest SNR.  The bottom panel of Fig. 8 shows a similar plot comparing the percent eclipse depths of the PED light curves to the percent eclipse depth of the integrated aperture light curves. Again, each object is represented by a single point corresponding to the quarter with the generated PED light curve that has the highest percent eclipse depth. A 1:1 reference line is plotted in green for both plots.

\begin{figure}[h!]
\center
\includegraphics[width=0.75\textwidth]{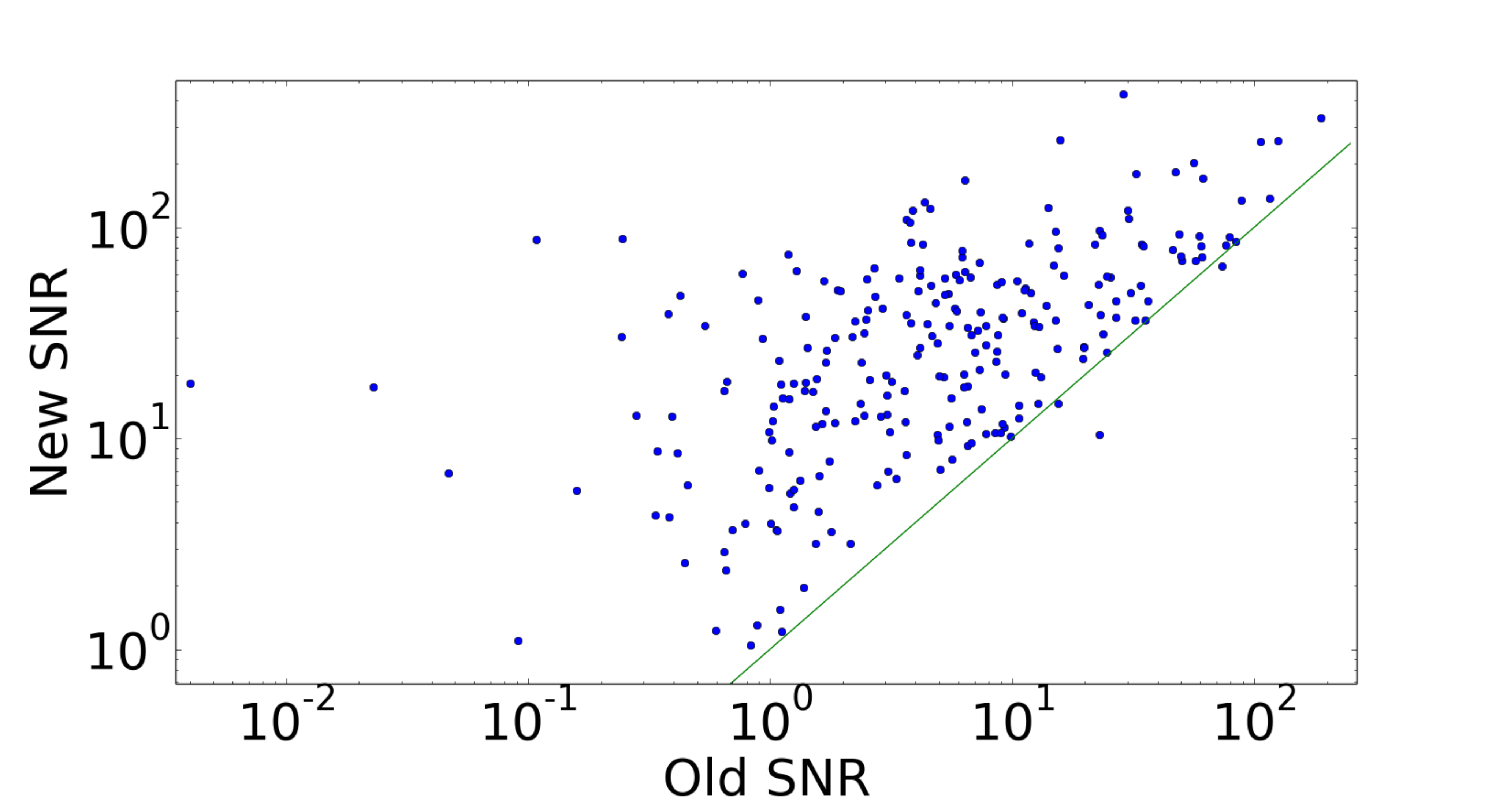}
\includegraphics[width=0.75\textwidth]{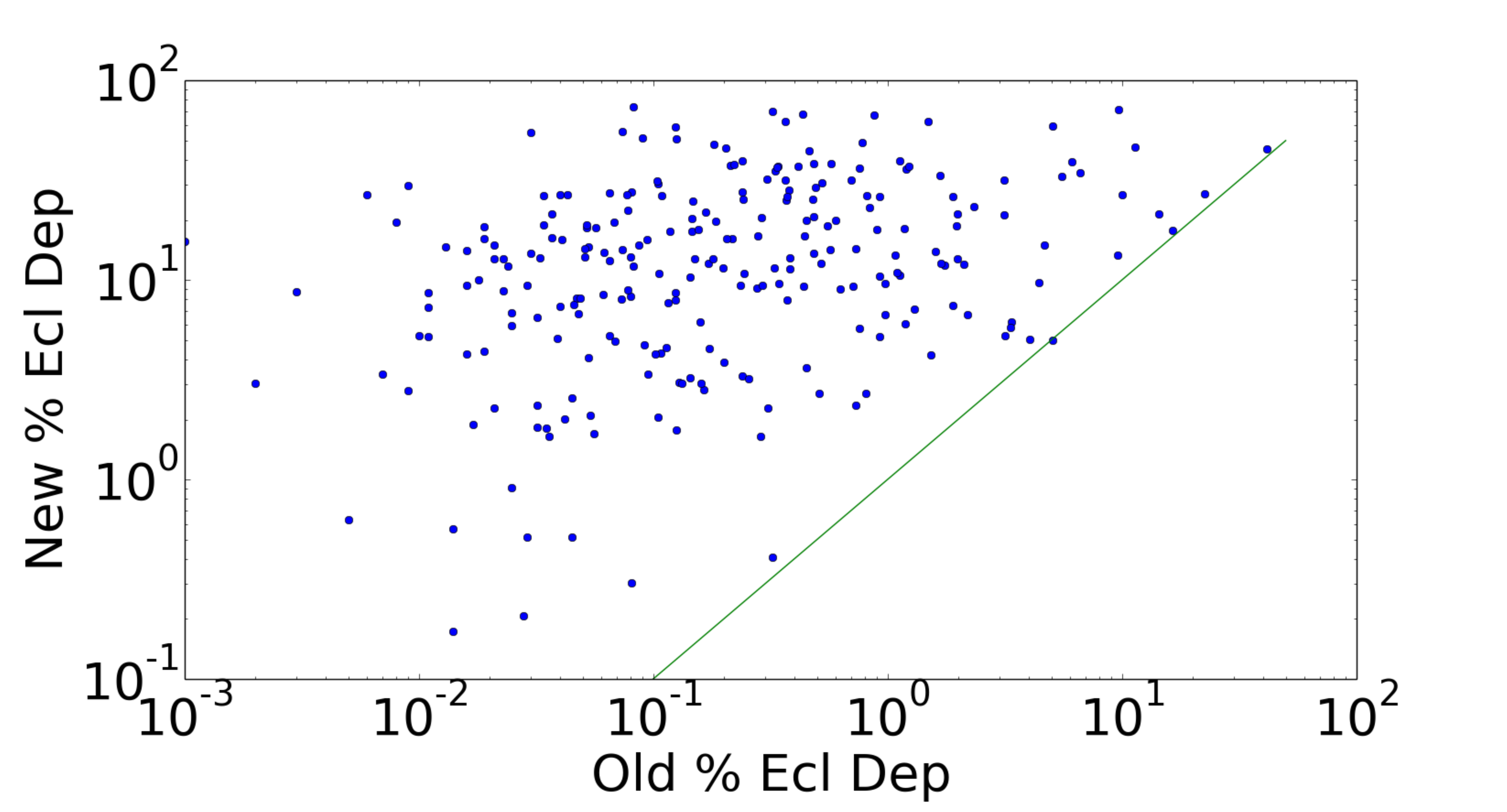}
\caption{
\label{fig:SNR_PED}
Top: the new signal-to-noise ratios of the SNR light curves versus the signal-to-noise ratios of the original light curves on a log-log scale.  Each object is represented by a single point, showing the quarter with the best new and original signal-to-noise ratio. Bottom: the new percent eclipse depths of the PED light curves versus the percent eclipse depths of the original light curves on a log-log scale.  Each object is represented by a single point showing the quarter with the best new and original percent eclipse depths. A 1:1 reference line is plotted in green in both plots. 
}
\end{figure}

To demonstrate the power of re-extraction methods, the original, SNR, PED and FED light curves for several representative examples are compared in Fig. 9.  Each plot is phase-folded and normalized. Each row shows the data from the same object with the first column representing the original light curve, the second representing the SNR light curve, the third representing the PED light curve and the fourth representing the FED light curve.  The objects chosen for presentation are representative of different regions in the signal-to-noise plot in the top panel of Fig. 8.

\begin{figure}[h!]
\center
\includegraphics[width=\textwidth]{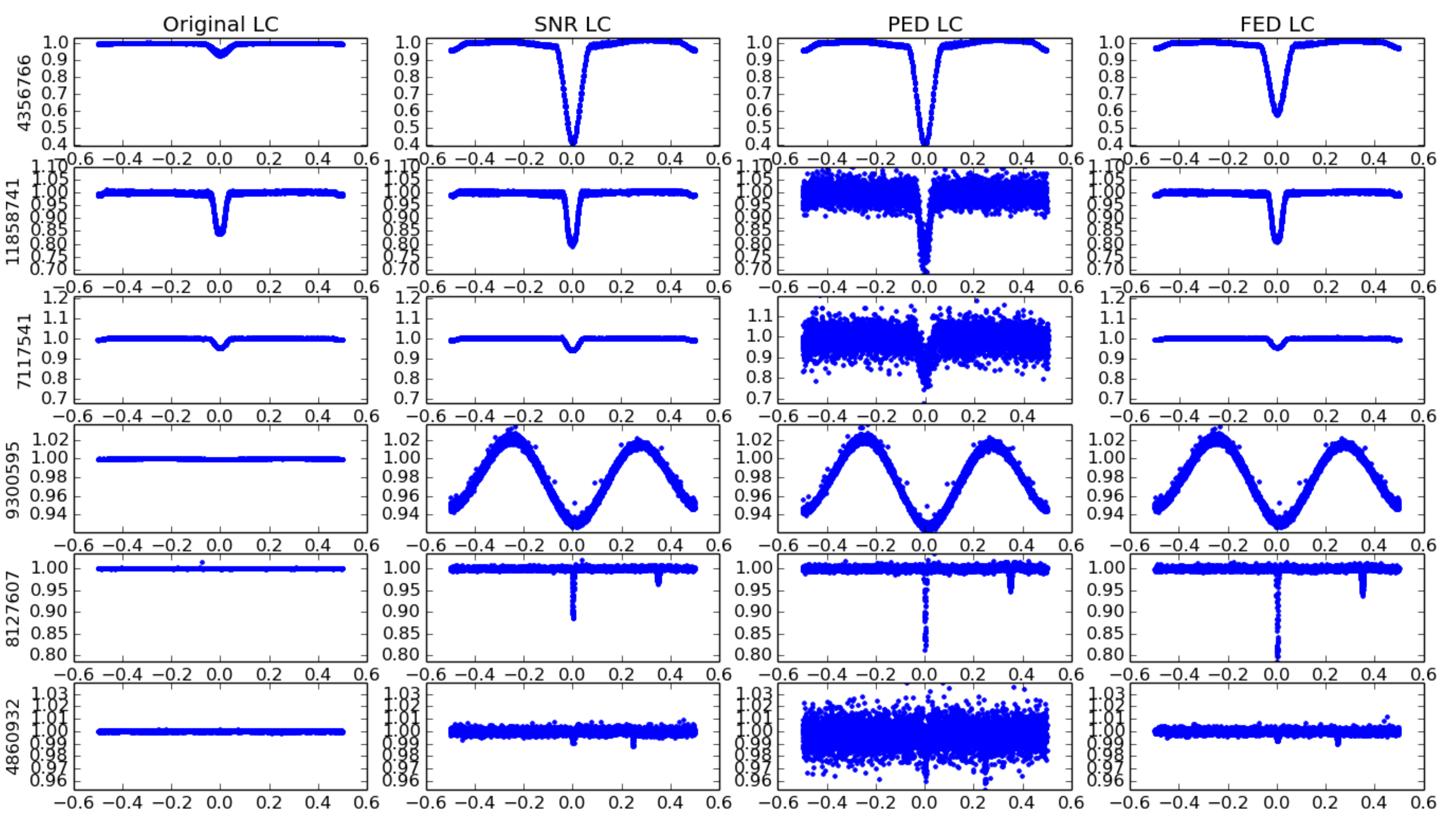}
\caption{
\label{fig:norm_phased}
Normalized phased comparison of the original and re-extracted light curves for several objects.  The rows represent different objects and for each row the y-axis is held constant.  Each column represents a different light curve where the first column contains the original, the second contains the SNR light curve, the third contains the PED light curve and the fourth contains the FED light curve. 
}
\end{figure}

\section{Discussion and Conclusions}
We have introduced a method for identifying \emph{Kepler} false positive eclipsing binaries that also can be used to identify the true source of the binary signal.  The pixel-LC tier is particularly useful for identifying which pixels in the TPF show the same signal as the original light curve. The pixel-pixel tier is quite efficient at showing where unique signals are localized. The SNR correlation and centroid tier is useful for determining which source is the true source of the eclipsing binary signal observed as well as identifying potential cross-talk events.  The SNR correlation tier differs from the correlation technique described in Bryson et al. (2013) in that we correlate the signals from each pixel to the signal from the original integrated aperture light curve instead of a model which assumes planetary transits.  This allows for more accurate correlation values for eclipsing binaries.  Ephemeris matching is useful for identifying target-target false positives and can also be used to identify cross-talk.  Our ephemeris matching technique is both similar and different from the method mentioned in Coughlin et al. (2014).  It is a combination of automated and manual steps while the previous method is completely automated.  Both techniques take into account the possibility of integer multiple periods and differing times of minimum, however our technique also accounts for primary and secondary eclipse swapping for cases where the primary and secondary eclipse depths are comparable in depth.  We also account for eclipse shape by manually analyzing ephemeris matches to ensure that objects with similar ephemerides also have the same phase-folded eclipse profiles.

The tiers in the false positive identification method have broader applications that can be applied to non-eclipsing binary objects and even to data obtained from other missions.  All three of these tiers can be used to find false positives for any type of object with a periodic signal, including variable stars and planets. These methods can also be used for other missions that use a similar pixel setup, such as the TESS mission (Sullivan et al. 2015).  While the TESS pixels are much larger than those of \emph{Kepler} (21`` versus 4``), the stars being observed are brighter and thus the profiles are very similar to those seen in \emph{Kepler} data.  These tiers can also be used to help refine K2 aperture sizes.  The pixel-pixel tier is particularly useful in this regard, as it can pick out unique signals without the need for a corresponding light curve, however this tier is not as effective as the SNR and centroid tier where the light curve is known. In addition to obtaining light curves for unobserved objects, these re-extraction methods can be used to further optimize the light curves of observed \emph{Kepler} targets as well.

We have shown that approximately 50\% of the identified false positive eclipsing binaries in the \emph{Kepler} FOV are target-object false positives and we have re-extracted new light curves for each. These new binaries have been added to the Catalog and their light curves have been made available. These re-extracted light curves show significant improvements in signal-to-noise ratios, percent eclipse depths and flux eclipse depths. Signals that were barely detectable before have become obvious, and percent eclipse depths have increased dramatically.  The average increase in SNR is $\sim27$ and the average increase in percent eclipse depth is $\sim15\%$. The percent eclipse depth was either improved or remained the same for every single re-extracted PED light curve when compared to the original light curve. For $\sim1\%$ of the re-extracted SNR light curves, we were not able to improve the signal-to-noise ratio from the original light curve. There are several factors that contribute to this. Due to the fact that \emph{Kepler} noise is not uncorrelated, it is difficult to predict the SNR of the final light curve from the SNRs of the individual pixels without combining them and recalculating the SNR. Table 1 lists a comparison of the original and re-extracted SNR, percent eclipse depth and flux eclipse depth for several objects.  A full version of Table 1 is available on the Catalog website.

\begin{table}[ht!]
\center{}
\begin{center}
\scriptsize \begin{tabular}{c c c c c c c c} 
\hline 
\hline 
FP KIC&\multicolumn{1}{p{1.8cm}}{\centering True KIC}&\multicolumn{1}{p{1.6cm}}{\centering Old SNR}&\multicolumn{1}{p{1.6cm}}{\centering New SNR}&\multicolumn{1}{p{1.8cm}}{\centering Old \% Ecl. Dep.}&\multicolumn{1}{p{1.8cm}}{\centering New \% Ecl. Dep.}&\multicolumn{1}{p{1.8cm}}{\centering Old Flux Ecl. Dep.}&\multicolumn{1}{p{1.8cm}}{\centering New Flux Ecl. Dep.}\\ 
\hline
3120742& 3120743 &12.221& 35.820& 0.748& 5.706& 160.578& 342.313\\
4861736& 4861747 &15.029& 25.897& 0.274& 8.757& 215.946& 326.920\\
5174959& 5174963 &25.479& 58.144& 3.150& 21.225& 257.198& 439.693\\
6309193& 6309195 &6.359& 62.513& 0.223& 16.215& 412.200& 684.438\\
6677267& 6677264 &0.108& 77.237& 0.008& 17.036& 3.469& 2681.716\\
8075755& 8075751 &15.499& 21.350& 1.804& 5.171& 166.220& 249.719\\
8879976& 8879975 &5.620& 15.190& 0.101& 4.330& 67.091& 157.405\\
9535881& 9535880 &4.361& 83.456& 0.206& 46.091& 123.304& 1281.006\\
10488450& 10488444 &2.796& 9.641& 0.036& 31.684& 69.560& 113.160\\
11756821& 11756823 &34.582& 42.104& 1.774& 8.995& 237.109& 467.455\\

\hline 
\end{tabular}
\caption{Sample table comparing the parameters from the original integrated aperture light curve to the new re-extracted parameters.  A full version of this table is available on the Catalog website.}
\label{tab:DV} 
\normalsize 
\end{center} 
\end{table}

The three new re-extracted light curves are beneficial in different ways.  The SNR light curve aims to minimize the noise and is most similar to the original integrated aperture light curve. Both maximize SNR but in different ways.  The original light curve maximizes total flux-to-noise while the SNR light curves maximize eclipse depth-to-noise.  The PED light curve determines the percent eclipse depth most accurately, however in doing so, the apertures tend to be more segmented and look less like the original integrated aperture light curves.  Furthermore, the PED light curves tend to be noisy and have the lowest SNR of the re-extracted light curves.  Despite the fact that the SNR and PED light curves always contain the same number of pixels, their apertures are rarely identical.  The FED light curve contains over 99\% of the signal from the binary that appears in the TPF and are most accurate at determining the flux eclipse depth.  For sources that are bright or on the edge of the aperture, there is a possibility that some of the pixels containing the binary signal were not included in the TPF and thus we are unable to get the signal contribution from these.  The same is true for the SNR and PED light curves, however, with the possibility that the pixels with the greatest SNR and percent eclipse depth may not be included in the TPF. While having the most accurate eclipse depth (as measured in units of flux) of the eclipsing binary source, the FED light curve suffers from dilution the most. Highly contaminated pixels with  low signal-to-noise ratios may still be included in the FED light curve if the flux eclipse depth of that pixel is high enough and the 99\% threshold has not been met.  This presents problems when trying to model the light curve as there is a significant amount of non-binary signal included in it.  This is not as big a problem for the other two re-extraction techniques, however it should still be considered.

To compare the light curve re-extraction method with the original extraction method, several well-isolated true source eclipsing binary targets with varying magnitudes were chosen and new light curves were re-extracted.  We found that, in general, the SNR light curve apertures tend to be smaller than their original aperture counterparts.  The disparity was greater for brighter sources and converged for dimmer sources.  For brighter sources, the FED light curve apertures were more consistent with the original apertures than the SNR light curve apertures, but for dimmer sources, the SNR aperture was more consistent with the original.  This is due to the differences in the signal-to-noise optimization between the SNR aperture and the original aperture.

We have demonstrated several ways to identify false positives and we have discussed methods for identifying true binary sources.  We have also demonstrated methods for re-extracting three new light curves that when considered together give a much more complete picture of the parameters of each new binary system.  We have shown that the newly extracted light curves improve several parameters vital to eclipsing binary statistics including both percent eclipse depth and flux eclipse depth. These new light curves are available on the Catalog website which is maintained at \texttt{http://keplerEBs.villanova.edu}.

In the future, certain aspects of our identification and re-extraction methods can be altered to improve individual light curves.  Improving the detrending algorithm can lead to reduced noise in the phase-folded and normalized light curves. This might lead to slightly higher SNR values, which can help ensure that pixels in apertures of low SNR cases are adjacent. The ephemeris matching technique we describe should be automated to reduce analysis time.  Our current re-extraction methods for background objects do not utilize the magnitude of the true source which can cause inappropriately sized apertures in rare cases, so incorporating the true source magnitudes for these objects can improve the SNR and percent eclipse depth further.

We thank Stephen Bryson for a helpful discussion and William Welsh for his valuable feedback.

We gratefully acknowledge support from NASA GO grant 14-K2GO1\_2-0057. DWL gratefully acknowledges support from the Kepler mission via NASA Cooperative Agreements NNX13AB58A and NNX11AB99A with the Smithsonian Astrophysical Observatory. The high-performance computational facility used for this work was sponsored in part by the Theodore Dunham, Jr. Grant of the Fund for Astrophysical Research. All of the data presented in this paper were obtained from the Multimission Archive at the Space Telescope Science Institute (MAST). STScI is operated by the Association of Universities for Research in Astronomy, Inc., under NASA contract NAS5-26555. Support for MAST for non-Hubble Space Telescope data is provided by the NASA Office of Space Science via grant NNX09AF08G and by other grants and contracts. Funding for this Discovery Mission is provided by NASA’s Science Mission Directorate. Facility: \emph{Kepler}.

\bibliographystyle{apj}

\end{document}